\documentclass[twocolumn,pra,showpacs,amsmath,amssymb]{revtex4}

\usepackage{amsmath,amsfonts,amsthm,graphicx,color}

\usepackage{float}
\usepackage{hyperref}

\begin{document}

\title{A two-qubit Bell inequality for which POVM measurements are relevant}
\author{T. V\'ertesi}
\email{tvertesi@dtp.atomki.hu}
\affiliation{Institute of Nuclear Research of the Hungarian Academy of Sciences\\
H-4001 Debrecen, P.O. Box 51, Hungary}
\author{E. Bene}
\email{bene@atomki.hu}
\affiliation{Institute of Nuclear Research of the Hungarian Academy of Sciences\\
H-4001 Debrecen, P.O. Box 51, Hungary}

\def\CC{\mathbb{C}}
\def\RR{\mathbb{R}}
\def\one{\leavevmode\hbox{\small1\normalsize\kern-.33em1}}
\newcommand*{\tr}{\mathsf{Tr}}
\newcommand*{\sgn}{sgn}
\newtheorem{theorem}{Theorem}
\newtheorem{lemma}[theorem]{Lemma}
\date{\today}

\begin{abstract}
A bipartite Bell inequality is derived which is maximally violated on the two-qubit state space if measurements describable by positive operator valued measure (POVM) elements are allowed rather than restricting the possible measurements to projective ones.
In particular, the presented Bell inequality requires POVMs in order to be maximally violated by a maximally entangled two-qubit state. This answers a question raised by N. Gisin.
\end{abstract}

\pacs{03.65.Ud, 03.67.-a}
\maketitle

\section{Introduction}\label{intro}
All pure bipartite entangled states violate the Clauser-Horne-Shimony-Holt (CHSH) Bell inequality \cite{Bell64,CHSH}, by performing appropriate measurements on the subsystem's state.  On the other hand, any Bell inequality which can be violated in the quantum world, can be maximally violated by some pure state and projective (von Neumann) measurements if no restrictions are put on the underlying Hilbert space. However, projective measurements are not the most general measurements. Still,  we have found no examples in the literature where general (so-called POVM) measurements would provide larger violation of some Bell inequalities by restricting to a given dimensional state space. In fact, this concerns a more specific problem posed recently by Gisin asking for a state for which POVM measurements would perform better than projective ones yielding larger violation of Bell inequalities \cite{Gisin07}.

In the present paper we study the above problems by restricting ourselves to the two-qubit space and to maximally entangled qubits, respectively. Note that in case of two-outcome Bell inequalities POVMs are not better than projective measurements  with respect to the amount of Bell violation \cite{WernerWolf,CHTW,Ito,LiangDoherty}. Further, in case of all those multiple-outcome Bell inequalities we are aware of in the literature, projective measurements still give maximal violation in the specific state space considered (see e.g. Refs~\cite{ADGL,NPA,BPAGMS,ZG,LLD}).

On a related note, we mention a recent result by Cabello and also by Nakamura on the Kochen-Specker theorem \cite{KS} proving that this theorem can be extended to a single qubit if POVM measurements can be used  instead of only projective ones \cite{Cabello}. Taking this result together with a method of Refs.~\cite{Aravind, CHTW} for building a $d\times d$ pseudo-telepathy game \cite{BBT05} from a $d$-dimensional Kochen-Specker construction, one may wonder whether this would entail a Bell inequality for a system of dimension $2\times 2$ where POVMs would give higher violation than projective ones. However, this approach turns out not to be feasible due to the proof of Brassard et al.~\cite{BMT}, stating that there is no pseudo-telepathy game of dimension $2\times 2$ even if POVMs are included.

Despite of the above negative results, we manage to derive a Bell inequality which proves that POVMs are relevant with respect to projective measurements for a two-qubit maximally entangled state and also for the case of a two-qubit space. First, Sec.~\ref{notation} introduces notation, then Sec.~\ref{povmqubit} reviews shortly the formalism of POVM and projective measurements on qubit states. In Sec.~\ref{3povm}, an optimization problem is presented considering positive definite matrices versus projection matrices. In Sec.~\ref{Bell}, a parametrized Bell inequality is given, for which the  quantum bound is calculated either including POVM measurements or restricting to projective measurements. We consider two cases, either the shared state is the two-qubit singlet state (Sec.~\ref{psiplus}) or it may be any state in the two-qubit state space (Sec.~\ref{twodim}). In Section~\ref{conc}, we conclude and pose open questions.

\section{Preliminaries}\label{notation}

Let us consider a standard Bell scenario \cite{Bell64}. Two
spacelike separated parties, Alice and Bob, share copies of a
quantum state $\rho$ in some dimension $d\times d$. The two
parties can choose among $N_A$ and $N_B$ different measurements
which are labelled by $x\in\{0,\ldots,N_A-1\}$ for Alice and
by $y\in\{0,\ldots,N_B-1\}$ for Bob, where we denote  the respective
outputs by $a\in\{0,\ldots,r_A-1\}$ and $b\in\{0,\ldots,r_B-1\}$.
In the most general description of a quantum measurement, $M_a^x$
($M_b^y$) denote the positive operator corresponding to outcome x
(y) when Alice (Bob) performs measurement a (b). Then the joint
conditional probabilities can be calculated in quantum theory by
the formula
\begin{equation}
p(ab|xy)=\tr(\rho M_a^x\otimes M_b^y). \label{pabxy}
\end{equation}
The positive operators above, summing to identity
$\sum_a{M_a^x}=\sum_b{M_b^y}=\one$, constitute POVMs for any inputs $x,y$.
However, in case of projection measurements, the positive operators $M_a^x$ and $M_b^y$ are projectors, hence the
ones belonging to the same inputs ought to be orthogonal to each other.

A Bell expression is a linear function $\vec
b\cdot\vec p=\sum_{a,b,x,y}b_{abxy}p(ab|xy)$ of the conditional probabilities $p(ab|xy)$ defined by Eq.~(\ref{pabxy}), where $\vec b$ has real components.
In order to maximize a Bell expression, it is enough to consider
pure states $\rho=|\psi\rangle\langle\psi|$. In our study, we will focus
on the state space of a pair of qubits, where up to a change of local
basis any pure state can be written as
$|\psi(\theta)\rangle=\cos(\theta)|00\rangle +
\sin(\theta)|11\rangle$. Now, let us take $\theta=\pi/4$, resulting in the maximally
entangled two-qubit state $|\psi^+\rangle=(|00\rangle+|11\rangle)/\sqrt 2$. Then, Gisin's problem in Ref.~\cite{Gisin07}  would be resolved by exhibiting a vector $\vec b$, where the maximum of $\vec b\cdot\vec p$ is achieved with POVM measurements (in contrast to the more restrictive case of projective measurements) on state $|\psi^+\rangle$.

In particular, we present a Bell expression (called $I_{CH3}$),
which consists of the CH expression,
\begin{align}
I_{CH}&\equiv -p_A(0|0)-p_B(0|0)+p(00|00)+p(00|01)\nonumber\\
&+p(00|10)-p(00|11),\label{ICH}
\end{align}
equivalent to the CHSH expression  \cite{CHSH}, and an expression involving on Alice's side a 3-outcome measurement ($x=2$)
\begin{align}
I_{3}&\equiv -p_A(0|2)-(1-1/\sqrt(2))p_A(1|2)+p(00|20)+p(00|21)\nonumber\\
&+p(10|20)-p(10|21), \label{I3}
\end{align}
resulting in the following Bell inequality,
\begin{equation}
I_{CH3}\equiv cI_{CH}+I_{3}\le 1, \label{ICH3}
\end{equation}
where $c>0$ is supposed. The local bound of 1 is obtained by examining all the
deterministic strategies with factorized joint probabilities
$p(ab|xy)=p_A(a|x)p_B(b|y)$, where $p_A(a|x)$, $p_B(b|y)$ denote local marginal
probabilities on Alice and Bob's respective sides. 

On one hand, we show by analytical means in Sec.~\ref{psiplus} that POVMs are required to obtain the
optimum value of $I_{CH3}$ on $|\psi^+\rangle$. On the other hand, it is shown 
(based on numerically exact computations) in Sec.~\ref{twodim} that, if we limit the
local dimension to two, it is still benefical to perform POVM measurements with respect to projective
ones. The above results have implications in the context of dimension witnesses as well \cite{dimwitpovm}.

\section{Measurements acting on qubits}\label{povmqubit}

A POVM is a family of positive operators $\{M_i\}$ with elements $M_i$, which sum to the identity, $\sum M_i=\one$.
In case of qubits, a Bell expression is optimized by pure states and extremal POVMs, whose elements $M_i$
are proportional to rank-1 projectors \cite{APP}.

For the case of binary outcomes and qubits, the extremal POVMs are projectors parametrized by a unit vector $\vec v=(v_x,v_y,v_z)$,
\begin{align}
M_0(\vec v)&=\frac{1}{2}(\one+\vec v\cdot\vec\sigma),\label{VN0}\\
M_1(\vec v)&=\frac{1}{2}(\one-\vec v\cdot\vec\sigma), \label{VN1}
\end{align}
where $\sigma=(\sigma_x,\sigma_y,\sigma_z)$ are the Pauli
matrices and the probability that outcome $i$ occurs for a state $\rho$ is given by Born's rule, $\tr(M_i(\vec v)\rho)$. Hence, for the maximally entangled state $\rho=|\psi^+\rangle\langle\psi^+|$, conditional probabilities~(\ref{pabxy}) read as
\begin{align}
p(ab=00|x(\vec a),y(\vec b))&=\langle\psi^+|M_{a=0}^{x(\vec a)} \otimes
M_{b=0}^{y(\vec b)}|\psi^+\rangle\nonumber\\&=1/4(1+\vec a'\cdot\vec b),
\label{VNjoint}
\end{align}
and the local marginal probabilities are given by
\begin{equation}
p_A(a=0|x(\vec a))=p_B(b=0|y(\vec b))=1/2. \label{VNmarg}
\end{equation}
In the above formulae, $x(\vec a)$ and $y(\vec b)$ label the
respective measurement settings of Alice and Bob parametrized by unit
vectors $\vec a, \vec b$ according to Eq.~(\ref{VN0}). Vector $\vec a'$ differs from $\vec a$
in a sign change of $a_y$, that is, $\vec a'=(a_x,-a_y,a_z)$.

For a three-outcome generalized POVM measurement acting on qubits, each of the three
extremal POVM elements $M_i$ is proportional to rank-1 projectors, hence we have
\begin{equation}
M_i=\lambda_i |w_i\rangle\langle w_i|,
\label{povm}
\end{equation}
with $\sum_{i=0}^2{M_i}=\one$, where $\lambda_i>0$ and $|w_i\rangle$  are normalized states.
On the other hand, for a three-outcome projective measurement on qubits, $M_i$ can be rank-$0,1,2$ projectors.
In case of rank 0 and rank 2, matrix $M_i$ is the zero and
identity matrix, respectively, whereas for rank 1, $M_i$ is defined by Eq.~(\ref{VN0}). Taking into account the
constraint that for qubits the sum of the ranks of $M_0$ and $M_1$ cannot
exceed 2, we have the following six possible pairs: $(0,0), (0,1), (1,0), (1,1), (0,2), (2,0)$, where the pair $(i,j)$ denotes the ranks of matrices $M_0$ and $M_1$, respectively.

\section{Case study}\label{3povm}

Let us consider the following optimization problem, which will
turn out to play a key role in the construction of the Bell
inequality $I_{CH3}$ defined by (\ref{ICH3}). First, let us define matrices
\begin{equation} F_0=\left(
\begin{array}{cc}
   1 &  0\\
   0 & -1\end{array} \right),
\label{F0}
\end{equation}
and
\begin{equation} F_1=\left(
\begin{array}{cc}
   1-\sqrt 2 &  1\\
   1 &  1-\sqrt 2\end{array} \right).
\label{F1}
\end{equation}
Then, we wish to maximize
\begin{equation}
W=\tr(M_0 F_0) + \tr(M_1 F_1),
\label{Wopt}
\end{equation}
over  $M_0, M_1, M_2$ positive $2\times 2$ matrices subject to  $M_0+M_1+M_2=\one$. Hence $M_i$, $i=0,1,2$ can be viewed as POVM elements of a three-outcome POVM. Let us denote by $\max{W_{POVM}}$ the maximum of (\ref{Wopt}) obtained in this way. Whereas, by further constraining $M_i$ to be projection matrices, we write the respective maximum as $\max{W_{proj}}$.
In the following subsections these values are given explicitly. 

\subsection{Maximum with POVMs}\label{maxpovm}

The optimization problem in Eq.~(\ref{Wopt}) with $M_i$ being POVM
elements is a typical instance of a semidefinite programming problem. Since
$F_0, F_1$ matrices are real valued, $\max{W_{POVM}}$ can be
obtained with real valued matrices $M_i$. By solving the SDP
problem using the package SeDuMi \cite{sedumi}, we obtain
\begin{equation}
w\equiv\max{W_{POVM}}=1.071~419~8987\ldots
\label{w}
\end{equation}
The matrices $M_i$ corresponding to this solution obey the positivity and unity conditions up
to high  precision ($\sim10^{-10}$). Below, they are written out using less digits,

\begin{equation}
M_0=\begin{pmatrix}
0.84153 &  -0.15627\\
-0.15627 & 0.02902
\end{pmatrix},
\label{M0}
\end{equation}

\begin{equation}
M_1=\begin{pmatrix}
0.14061 &  0.25242\\
0.25242 & 0.45314
\end{pmatrix},
\label{M1}
\end{equation}
and by definition $M_2=\one-M_0-M_1$. As it can be checked, each of these truncated matrices has positive eigenvalues, hence they define a valid POVM. By substituting these matrices into the expression $W$ in  (\ref{Wopt}), we regain the value of $w$ in (\ref{w}) up to 5 digits. 

\subsection{Maximum with projection matrices}\label{maxproj}

Now, let $M_i$  be two-dimensional projection matrices in the optimization problem~(\ref{Wopt}). According to the ranks of $(M_0,M_1)$, we have the six possibilities, $(0,0), (0,1), (1,0), (1,1), (0,2), (2,0)$, listed in Sec.~\ref{povmqubit}.
The corresponding $W_{proj}$ values read as follows (using the parametrization of rank-1 projectors in Eqs.~(\ref{VN0},\ref{VN1})),

\begin{center}
\begin{tabular}{cc}
Ranks & $W_{proj}$\\
\hline\hline
   $(0,0)$ &  $0$\\
   $(0,1)$ &  $-v_x+1-\sqrt 2$\\
   $(1,0)$ &  $v_z$\\
   $(1,1)$ &  $-v_x+v_z+1-\sqrt 2$\\
   $(0,2)$ &  $2-2\sqrt 2$\\
   $(2,0)$ &  $0$
\end{tabular}
\end{center}
Since $\vec v$ is a unit vector, $\max\{-v_x+v_z\}=\sqrt 2$, and according to the table, we obtain $\max{W_{proj}}=1$. Comparing this value with the value of $w$ in (\ref{w}), shows that indeed POVM elements give benefit over projectors in the presented optimization problem.

\section{Bell expression $I_{CH3}$}\label{Bell}

We next try to explain the construction of  the Bell expression $I_{CH3}$ (introduced under (\ref{ICH3})) building upon the results of the optimization problem in the previous section. In particular, we wish to achieve somehow that matrices $F_0$ and $F_1$ defined by  Eqs.~(\ref{F0},\ref{F1}) would naturally arise in a Bell scenario.
For this sake, suppose that Alice shares with Bob the maximally
entangled quantum state $|\psi^+\rangle$ and the optimal POVM
elements $M_{0,1,2}$ in Eqs.~(\ref{M0},\ref{M1}) correspond to Alice's
3-outcome measurement, $M_{a}^{x=2}=M_a$ for $a=0,1,2$. Moreover, let us assume that Bob has two binary-outcome settings, where the measurement operators corresponding to outcome 0 are described by the following projectors,
\begin{align}
M_{b=0}^{y=0}=\frac{1}{2}(\one+\vec b_0\cdot\vec\sigma),\nonumber\\
M_{b=0}^{y=1}=\frac{1}{2}(\one+\vec b_1\cdot\vec\sigma), \label{Mb01}
\end{align}
with $\vec b_0=(1/\sqrt2,0,1/\sqrt2)$ and $\vec
b_1=(-1/\sqrt2,0,1/\sqrt2)$. Then $F_0$ and $F_1$ matrices of Eqs.~(\ref{F0},\ref{F1}) can be reproduced in the following way,
\begin{align}
F_0/\sqrt 2&=-\one+M_{b=0}^{y=0}+M_{b=0}^{y=1}\nonumber\\
F_1/\sqrt 2&=-(1-1/\sqrt 2)\one+M_{b=0}^{y=0}-M_{b=0}^{y=1},
\label{F01}
\end{align}

For a maximally entangled state $|\psi^+\rangle$ and for real valued $2\times 2$ matrices
$A,B$, we have the expectation value $\langle \psi^+|A\otimes
B|\psi^+\rangle=\tr(AB)/2$. Hence, due to equations in (\ref{F01}) the optimization problem~(\ref{Wopt}) can be
seen as maximizing the expression $2\sqrt 2 I_3$  of Eq.~(\ref{I3}) over Alice's measurement $\{M_0=M_{a=0}^{x=2},M_1=M_{a=1}^{x=2}\}$ 
assuming Bob's projectors $M_{b=0}^{y=0,1}$ are defined by (\ref{Mb01}).

Let us now consider the Bell inequality $I_{CH3}$ of (\ref{ICH3}),
\begin{equation}
I_{CH3}=cI_{CH}+I_{3}\le 1,
\end{equation}
with $c$ positive. In particular, if $c$ is very large, then the CH expression~(\ref{ICH}) becomes dominant in
$I_{CH3}$, entailing that the maximum quantum violation can
be obtained by projection operators
$M_{a=0,1}^{x=0},M_{b=0,1}^{y=0}$ very close to the ones which
maximize the CH expression. Note that for $I_{CH}$ the maximum quantum value of 
 $(\sqrt2-1)/2$ can be obtained by the state
$|\psi^+\rangle$, and by projection operators $M_{b=0}^{y=0,1}$ defined by (\ref{Mb01}) on Bob's side.
Assuming the above ideal case for Bob's operators
and a state $|\psi^+\rangle$, we obtain that $I_{CH3}$ is maximal
if Alice's three-outcome measurement ($x=2$) consists of POVM elements~(\ref{M0},\ref{M1}) resulting from the
optimization problem~(\ref{Wopt}). On the other side, it is
expected that if $c$ is not extreme large, then Bob's optimal
operators $M_{b=0}^{y=0,1}$ would differ somewhat from the ideal
CH-violating ones, but should still be close to it, so that
Alice's three-outcome measurement ($M_a^{x=2}$) would still prefer POVM
elements with respect to projection-valued elements in order to
get maximum violation of $I_{CH3}$. In the following, the validity
of the above reasoning will be supported by explicit calculations.
First, in Sec.~\ref{lower}, a lower bound is established on the violation of $I_{CH3}$
applying POVMs on $|\psi^+\rangle$. In Sec.~\ref{Bellproj}, Bell inequalities are derived from $I_{CH3}$
restricting to projective measurements on the two-qubit space.
Then, in Sec.~\ref{psiplus} and in Sec.~\ref{twodim}, the maximal violation of inequality $I_{CH3}$ is calculated in 
case of projective measurements acting on $|\psi^+\rangle$ and on the two-qubit state space, respectively. In both cases a strictly smaller
violation of $I_{CH3}$ than with the use of POVM measurements is found.

\subsection{Lower bound with POVM}\label{lower}

A useful lower bound on $I_{CH3}$ can be obtained for a pair of
maximally entangled qubits, and for any two-qubit state as well in
the following manner. Let us take as a special choice the state $|\psi^+\rangle$ and
those measurement operators which maximize $cI_{CH}$ in (\ref{ICH3}), giving the value of
$c(\sqrt2-1)/2$. However, with these operators for Bob, as
discussed earlier, we have the maximum value of $w/(2\sqrt2)$ for
$I_{3}$, where $w$ comes from~(\ref{w}). Adding up the two values
according to Eq.~(\ref{ICH3}), we have the following lower bound on the expression $I_{CH3}$,
\begin{equation}
\max_{POVM,\psi^+}I_{CH3}\ge c(\sqrt2-1)/2 + w/(2\sqrt2)
\label{ICH3povm}
\end{equation}
for POVM measurements acting on the state $|\psi^+\rangle$ and on any two-qubit state as well.
We wish to mention that any extremal POVM measurement with real coefficients on the qubit space
in the form of~(\ref{povm}) can be reproduced with rank-1 von Neumann
measurements with real coefficients acting on the qutrit space.
This is due to Neumark's theorem \cite{Neumark},
stating that POVM measurements can always be seen as projective
measurements acting on a larger Hilbert space. For a three-outcome measurement, an explicit construction is given by \cite{POVMviaVN}.

\subsection{Deriving Bell inequalities in case of projective measurements on qubits}\label{Bellproj}

Here we study the maximum quantum violation of $I_{CH3}$ if only projective
measurements are allowed on the local qubit spaces. First, let us note that in
order to violate any two-party Bell inequality, each party
must have at least two non-degenerate operators belonging to
different measurement settings, otherwise the quantum predictions
could be simulated within a local classical model. Concerning a pair
of qubits and inequality $I_{CH3}$, this entails that both of Bob's
operators must be rank-1 projectors. We now state the following
lemma:
\begin{lemma}\label{lemma1}
Consider inequality $I_{CH3}$ and assume that only projective measurements can be performed on the local qubit spaces. In this case, $I_{CH3}$ can be violated only if Alice's and Bob's measurement operators corresponding to $I_{CH}$ are rank-1 projectors.
\end{lemma}
The proof of this lemma can be found in \cite{proof1}. Since we are interested in the non-trivial case that $I_{CH3}$ can be violated, due to the above lemma, Alice's two operators $M_{a=0}^{x=0,1}$ can be considered as rank-1 projectors. Then, we are left with six possible cases, $(0,0), (0,1), (1,0), (1,1), (0,2), (2,0)$ according to the ranks of $M_{a=0}^{x=2}$ and $M_{a=1}^{x=2}$, as it was discussed in Sec.~\ref{povmqubit}. 
In each case above the original inequality $I_{CH3}$ is modified as follows. If a measurement $M_a^x$ is rank 0, 
we set $p_A(a|x)=0$, $p(ab|xy)=0$. In case that $M_a^x$ is rank 2, we set $p_B(b|y)=1$,
$p(ab|xy)=p(b|y)$. On the other hand, if
both $M_{a=0}^{x=2}$ and $M_{a=1}^{x=2}$ are rank-1 projectors, we have
$p_A(0|2)+p_A(1|2)=1$ and $p(00|xy)+p(10|xy)=p_B(0|y)$ for $y=0,1$. In this way we derive two-outcome Bell
inequalities from $I_{CH3}$, which look as follows,
\begin{align}
I_{00}\equiv&cI_{CH}\le 0 \label{I00}\\
I_{01}\equiv&cI_{CH}+p_A(0|2)+p(00|20)-p(00|21)\le 1/\sqrt 2 \label{I01}\\
I_{10}\equiv&cI_{CH}-p_A(0|0)+p(00|00)+p(00|01)\le 1 \label{I10}\\
I_{11}\equiv&cI_{CH}-1/\sqrt 2p_A(0|2)+p_B(0|0)-p_B(0|1)\nonumber\\
&+2p(00|21)+1/\sqrt 2 -1\le 1 \label{I11}\\
I_{02}\equiv&cI_{CH}+p_B(0|0)-p_B(0|1)\le 1 \label{I02}\\
I_{20}\equiv&cI_{CH}+p_B(0|0)+p_B(0|1)-1\le 1 \label{I20}
\end{align}
where $I_{ij}$ denotes Bell expression derived from $I_{CH3}$ by setting $M_{a=0}^{x=2}$
to be rank-$i$ projector, and $M_{a=1}^{x=2}$ to be rank-$j$ projector.
So, in order to get the maximum violation of $I_{CH3}$ by von Neumann's projective
measurements on qubits, we are left with calculating maximum quantum violation of
the above inequalities~(\ref{I00}-\ref{I20}) by considering rank-1 projectors. This is just what we will do in the following by considering maximally entangled qubits (Sec.~\ref{psiplus}) and also the state space of two qubits (Sec.~\ref{twodim}).

\subsection{Maximizing $I_{CH3}$ with projective measurements on maximally entangled qubits}\label{psiplus}

First, note that expressions $I_{20}$ and $I_{02}$ are equivalent up
to relabelling of the outcomes. Further, for $|\psi^+\rangle$,
expressions $I_{00}$, $I_{02}$, $I_{20}$ coincide giving the quantum maximum of
$c/2(\sqrt 2 -1)$. In order to calculate the quantum maximum for the remaining three cases $I_{01}$, $I_{10}$, $I_{11}$, we present the following lemma:
\begin{lemma}\label{lemma2} 
Let us assume that $\vec b_i$, $i=1,2$ are unit vectors in the Euclidean space. Then we have
$\max_{\vec b_1, \vec b_2}\left\{\left|\vec b_1 + \vec
b_2\right| + k\left|\vec b_1 - \vec
b_2\right|\right\}=2\sqrt{1+k^2}$.
\end{lemma}

Using this lemma, Eqs.~(\ref{VNjoint},\ref{VNmarg}), and the simple fact that $\vec a\cdot\vec c\le \left| \vec c \right|$ for a unit vector $\vec a$, we obtain the following quantum maximum for
the Bell expressions~(\ref{I00}-\ref{I20}) with rank-1 projective measurements on $|\psi^+\rangle$:

\begin{center}
\begin{tabular}{cc}
Expression & Maximum on state $|\psi^+\rangle$\\
\hline\hline
   $I_{00},I_{02},I_{20}$ &  $\frac{1}{2}c(\sqrt 2 -1)$\\
   $I_{01}$ &  $\frac{1}{2}(1/\sqrt 2 -1-c+\sqrt{c^2+(c+1)^2}$\\
   $I_{10}$ &  $\frac{1}{2}(-c+\sqrt{c^2+(c+1)^2})$\\
   $I_{11}$ &  $\frac{1}{2}(1/\sqrt 2+c(\sqrt2-1)).$
\end{tabular}
\end{center}
We can observe the simple relations $I_{00}<I_{11}$ and $I_{01}<I_{10}$ between
the right-hand side formulae. On the other hand, using the
relationship between the quadratic and arithmetic mean, we have
$I_{10}>I_{11}$ for $c>0$. Thus for $c>0$, the quantum maximum  of expression $I_{CH3}$ with projective measurements on $|\psi^+\rangle$, 
is provided by expression $I_{10}$, yielding the value of
\begin{equation}
\max_{proj,\psi^+}I_{CH3}=\frac{-c+\sqrt{c^2+(c+1)^2}}{2}.
\label{ICH3proj}
\end{equation}
On the other hand, we have the lower bound ~(\ref{ICH3povm}) of expression $I_{CH3}$ with POVM measurements on $|\psi^+\rangle$,
where $w$ is defined by (\ref{w}). Note that Eq.~(\ref{ICH3proj}) becomes bigger than 1 (i.e., the local
bound on $I_{CH3}$) for $c>3$. Hence, in the following only the interval $c>3$ will be considered. By equating the right-hand side of Eqs.~(\ref{ICH3povm}) and (\ref{ICH3proj}), and solving the equation for
$c$, we obtain the value of $(2-w^2)/(4w-4)\simeq2.9826$. 
Hence, for $c>3$, the right-hand side of Eq.~(\ref{ICH3povm}) is definitely bigger than of Eq.~(\ref{ICH3proj}).
This implies that for $c>3$, Bell inequality $I_{CH3}$ in (\ref{ICH3}) is stronger violated by POVMs on a pair of maximally entangled qubits  than by considering only projective measurements. This answers Gisin's question \cite{Gisin07} in the positive.

\begin{figure}
\includegraphics[width=\columnwidth]{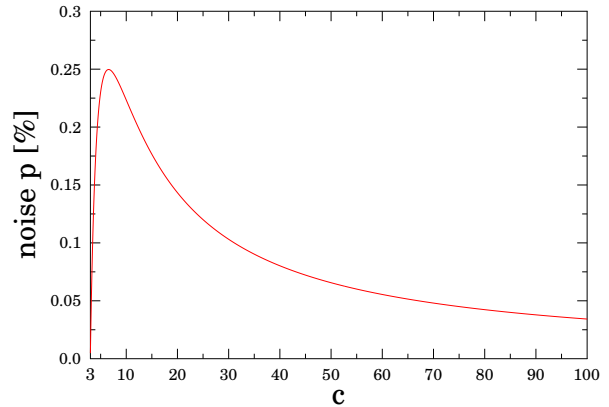} \caption{
Noise thresholds in function of parameter $c$ such that POVMs on a
two-qubit Werner state \cite{Werner} with mixed noise $p$ ceases
to be better than projective measurements. The curve gives an analytical lower bound for $p$.} \label{noise}
\end{figure}

We may have a quantitative measure about the performance of POVMs
over projective measurements by adding a fraction of $p$ white
noise to the maximally entangled two-qubit state \cite{Werner},
\begin{equation}
\rho(p)=(1-p)|\psi^+\rangle\langle\psi^+|+p\one/4,
\label{rhop}
\end{equation}
such that POVMs on state $\rho(p)$ still perform better than projective
ones. In case of $\rho(p)$, a lower bound on the violation of
$I_{CH3}$ for POVM measurements is given by $(1-p)\max_{POVM,\psi^+}I_{CH3}+(-2c+2(1/\sqrt 2 -1)\tr{M_1})p/4$, with $M_1$ defined in Eq.~(\ref{M1}).
Fig.~\ref{noise} shows in function of $c$ the amount of white noise which can be
tolerated due to the above formula such that POVM measurements are still better than
projective ones. The lower bound of $p=0.249717\%$ on the maximum tolerable noise is attained
by $c\simeq 6.56182$.

\subsection{Maximizing $I_{CH3}$ with von Neumann measurements on a pair of qubits}\label{twodim}

We now fix $c=100$ and show that if the state is allowed to be any two-qubit
state, POVMs can still perform better than projectors. Our task is
to compute the maximum quantum violation for the derived
Bell inequalities (\ref{I00}-\ref{I20}) assuming that rank-1 projectors act
on the two-qubit space. However, we can establish an upper bound
on these values by computing the SDP hierarchy of Navascu\'es, Pironio and Ac\'in \cite{NPA} on various levels. Incidentally, the lower bound achievable with real-valued rank-1 projectors on qubits coincide with the upper bound value coming
from the SDP calculation of \cite{NPA} on level two for each inequality in the set~(\ref{I00}-\ref{I20}). We collected in the table below the obtained values. Note that Bell expressions $I_{02}$ in Eq.~(\ref{I02}) and $I_{20}$ in Eq.~(\ref{I20}) are equivalent to each other.

\begin{center}
\begin{tabular}{cc}
Expression & Maximum on qubits\\
\hline\hline
   $I_{00}$ &      20.71068 \\
   $I_{01}$ &      20.91928\\
   $I_{10}$ &      21.06690\\
   $I_{11}$ &      21.06801\\
   $I_{02},I_{20}$ &   20.71775
\end{tabular}
\end{center}
This table shows that by considering projective measurements the two-qubit maximum of expression $I_{CH3}$ for $c=100$
is provided by expression $I_{11}$, yielding numerically the value of $21.06801$, whereas owing to Eq.~(\ref{ICH3povm}) the lower bound on the two-qubit maximum by considering POVMs is $100(\sqrt(2)-1)/2 + w/2\simeq
21.0895$. This proves the existence of bipartite Bell inequalities for which maximal violation on
the two-qubit space can be achieved only with the use of generalized POVM measurements.

Note that any bipartite Bell inequality consisting of two measurement settings with two outcomes each on Bob's side can be maximally violated on the two-qubit space. This follows from the works of Refs.~\cite{Masanes,TV06}. In particular, due to Lemma~2 of Ref.~\cite{TV06}, the quantum maximum is achieved by a state with support on Bob's qubit. However, using the Schmidt decomposition theorem, it induces the composite space to be a pair of qubits, letting Alice's state space be a qubit as well. In light of this, $I_{CH3}$ is a Bell inequality whose maximal violation is attained by performing POVM measurements on qubits. On the other hand, if only projective measurements are allowed, then qutrits are needed to achieve maximal violation.

\section{Conclusion}\label{conc}
Though there are indications that performing POVM measurements on
a given state or on a given state space may yield benefit over projective ones, the question
has not been settled yet. In the present paper we provide a
bipartite Bell inequality with a small number of inputs and
outputs which answers this question in the positive. Moreover, we
found that the improvement, which we defined in terms of noise
resistance is not marginal. It may be within the range of what is
feasible experimentally nowadays. 

However, one may still wonder
whether it would be possible to construct even better Bell inequalities with
more settings or with more parties allowing a
bigger separation in the maximum of Bell values achievable with POVM versus
projective measurements on a given state. A bigger gap might be suggested by the amount
of communication to simulate different types of measurements on a
singlet state. Whereas for projective measurements one bit of
communication suffices \cite{TonerBacon}, for POVM measurements
the best protocol constructed so far needs on average six bits of
communication \cite{Methot}. Based on the best local models
constructed for POVMs and for projective measurements on a mixture
of $d$-dimensional maximally entangled states with noise \cite{APBTA}, it
is also plausible that moving from qubits to higher dimensions,
POVM measurements become much more efficient than projective ones. 

\acknowledgments We would like to thank Jean-Daniel Bancal, Nicolas Brunner, Stefano Pironio
and our colleague K\'aroly F. P\'al for useful discussions. T.V. has been supported by a J\'anos Bolyai
Programme of the Hungarian Academy of Sciences.


\begin{thebibliography}{99}
\bibitem{dimwitpovm}
According to the definition introduced in Ref.~\cite{BPAGMS}, a two-dimensional witness is a vector
$\vec w$ of real coefficients, such that $\vec w\cdot \vec p=\sum_{a,b,x,y}w_{abxy}p(ab|xy)\le
w_2$ for all probabilities $p(ab|xy)$ expressed by formula~(\ref{pabxy})
with a two-qubit state $\rho$, such that there exist quantum
correlations beyond local dimension two for which $\vec w\cdot\vec
p>w_2$. In light of our result, in order to sensibly define the
constant $w_2$ for any $\vec w$, it is not enough to carry out
projective measurements in the two-qubit space, but the inclusion
of the most general POVM measurements are needed too.
References \cite{BPAGMS,WCD,BBT,VP09,WP,Junge} provide various constructions of dimension witnesses.
\bibitem{POVMviaVN}
Let us use Eq.~(\ref{povm}), and write $|w_i\rangle$ as real-valued
normalized qubit states, $|w_i\rangle=w_{i0}|0\rangle +
w_{i1}|1\rangle$. Let us write rank-1 projectors in the qutrit
space as $P_i=|w'_i\rangle\langle w'_i|$, where
$|w'_i\rangle=\sqrt{\lambda_i}w_{i0}|0\rangle +
\sqrt{\lambda_i}w_{i1}|1\rangle +
s_{jk}\sqrt{1-\lambda_i}|2\rangle$. Here $i\neq j\neq k$ and
$s_{jk}=-\sgn{(\langle w_j|w_k\rangle)}$. It can be checked that
in this way $P_i$ satisfies the relation
$P_iP_j=\delta_{ij}P_i$, thus $P_i$ are rank-1 projectors corresponding
to a von Neumann measurement. Moreover, for any state $\rho$ in the
qubit space, we get the equality of probabilities, $\tr(M_i\rho)=\tr(P_i\rho)$, where $M_i$ are defined by Eq.~(\ref{povm}).
\bibitem{proof1}
Let us calculate separately the
quantum maximum for expression $I_{CH}$ and $I_{3}$ assuming at least one
degenerate operator (i.e., rank-0 or rank-2 projectors) among $M_{a=0}^{x=0,1}$ and $M_{b=0}^{y=0,1}$. 
The $I_{3}$ part contains on Alice's side only one setting, which means that the quantum maximum cannot be bigger than the local maximum of $1$. The $I_{CH}$ part gives at most zero if one of the settings is degenerate. This implies the upper
bound value of $1+0=1$ on the quantum maximum for expression $I_{CH3}$ if one of
Alice's operators $M_{a=0}^{x=0,1}$ or Bob's operators $M_{b=0}^{y=0,1}$ is degenerate. This entails Lemma~\ref{lemma1}.
\end{thebibliography}
\end{document}